\documentclass[notitlepage,nofootinbib,preprintnumbers,amssymb,superscriptaddress]{revtex4-1}
\usepackage{amsfonts,amssymb,mathtools,graphicx,color,bm}
\definecolor{ultramarine}{rgb}{0.07, 0.04, 0.56}
\definecolor{cadmiumgreen}{rgb}{0.0, 0.42, 0.24}
\definecolor{indigo(dye)}{rgb}{0.0, 0.25, 0.42}
\usepackage[linktocpage=true]{hyperref}
\usepackage{here}
\hypersetup{
colorlinks=true,
citecolor=ultramarine,
linkcolor=cadmiumgreen,
urlcolor=indigo(dye),
}

\usepackage{autobreak}
\usepackage{subfigure}
\newcommand{\fr}[2]{\frac{#1}{#2}}
\newcommand{\D}{{\rm{d}}}
\newcommand{\E}{{\rm{e}}}
\newcommand{\pa}{\partial}

\newcommand{\na}{\nabla}
\newcommand{\bra}[1]{\left( #1 \right)}  
\newcommand{\brb}[1]{\left[ #1 \right]}  
  
\newcommand{\be}{\begin{equation}}  
\newcommand{\ee}{\end{equation}}
\newcommand{\bem}{\begin{bmatrix}}
\newcommand{\eem}{\end{bmatrix}}
\newcommand{\Mpl}{M_{\rm Pl}}

\newcommand{\ep}{\epsilon}

\newcommand{\la}{\lambda}
\newcommand{\si}{\sigma}

\usepackage{wrapfig}
\usepackage{datetime}

\allowdisplaybreaks

\begin{document}

\preprint{KOBE-COSMO-21-14, YITP-21-56}

\title{Cosmic No-hair Conjecture and Inflation with an SU(3) Gauge Field}

\author{Pengyuan Gao}
\affiliation{Department of Physics, Kobe University, Kobe 657-8501, Japan}

\author{Kazufumi Takahashi}
\affiliation{Center for Gravitational Physics, Yukawa Institute for Theoretical Physics, Kyoto University, Kyoto 606-8502, Japan}

\author{Asuka Ito}
\affiliation{Department of Physics, Tokyo Institute of Technology, Tokyo 152-8551, Japan}
\affiliation{Department of Physics, National Tsing-Hua University, Hsinchu 30013, Taiwan}
\affiliation{Center for Theory and Computation, National Tsing-Hua University, Hsinchu 30013, Taiwan}

\author{Jiro Soda}
\affiliation{Department of Physics, Kobe University, Kobe 657-8501, Japan}

\begin{abstract}

We study inflationary universes with an SU(3) gauge field coupled to an inflaton through a gauge kinetic function. 
Although the SU(3) gauge field grows at the initial stage of inflation due to the interaction with the inflaton, nonlinear self-couplings in the kinetic term of the gauge field become
significant and cause nontrivial dynamics after sufficient growth.
We investigate the evolution of the SU(3) gauge field numerically and reveal attractor solutions in the Bianchi type I spacetime. 
In general cases where all the components of the SU(3) gauge field have the same magnitude initially, they all tend to decay eventually because of the nonlinear self-couplings.
Therefore, the cosmic no-hair conjecture generically holds in a mathematical sense. 
Practically,  however, the anisotropy can be generated transiently in the early universe. 
Moreover, we find  particular cases for which several components of the SU(3) gauge field survive against the nonlinear self-couplings.
It occurs due to flat directions in the potential of a gauge field for Lie groups whose rank is higher than one.
Thus, an SU(2) gauge field has a specialty among general non-Abelian gauge fields. 
\end{abstract}
\maketitle

\tableofcontents

\section{Introduction}
No matters can survive in an expanding homogeneous universe in the presence of a positive cosmological constant except for the Bianchi type \mbox{I\hspace{-1pt}X} spacetime~\cite{Wald:1983ky}.
Due to this cosmic no-hair theorem, it is believed that  a hair such as a vector field never survives during inflation because an inflaton mimics the role of the cosmological constant. 
It is often called the cosmic no-hair conjecture.
Although there have been several attempts to seek for a counterexample to the conjecture~\cite{Ford:1989me,Golovnev:2008cf,Kanno:2008gn,Ackerman:2007nb}, they suffer from instabilities in the models~\cite{Himmetoglu:2008zp,EspositoFarese:2009aj}.
However, a healthy counterexample to the conjecture motivated by supergravity was found in \cite{Watanabe:2009ct}, where a vector field is coupled to an inflaton through a gauge kinetic function.  
The point of the model is that the inflaton does not mimic a positive cosmological constant exactly, whose deviation is characterized by the slow-roll parameter.
Then, an inflationary universe with a small anisotropy proportional to the slow-roll parameter can be realized~\cite{Watanabe:2009ct,Soda:2012zm,Maleknejad:2012fw,Maleknejad:2012as}.

Importantly, anisotropic inflation yields several observational signatures such as statistical anisotropy~\cite{Gumrukcuoglu:2010yc,Dulaney:2010sq,Watanabe:2010fh,Watanabe:2010bu,Hervik:2011xm,Thorsrud:2012mu,Bartolo:2012sd,Bartolo:2011ee,Abolhasani:2013zya,Ohashi:2013qba,Shiraishi:2013oqa,Chen:2014eua,Naruko:2014bxa,Emami:2015uva}.
They have been tested by the observations of the cosmic microwave background~\cite{Akrami:2018odb,Ramazanov:2016gjl} and the large-scale structure of the Universe~\cite{Sugiyama:2017ggb}.
Implications from the observations~\cite{Fujita:2017lfu,Talebian:2019opf} and 
future perspectives are discussed~\cite{Shiraishi:2016omb,Bull:2018lat}.
Considering phenomenological and observational importance, 
it is worth extending the anisotropic inflation model as far as possible~\cite{Kanno:2009ei,Kanno:2010nr,Emami:2010rm,Yamamoto:2012tq,Yamamoto:2012sq,Murata:2011wv,Maeda:2012eg,Maeda:2013daa,Do:2011zza,Do:2011zz,Ohashi:2013pca,Ohashi:2013mka,Bartolo:2013msa,Emami:2013bk,Do:2013tea,Chen:2014zoa,Bartolo:2014hwa,Ito:2015sxj,Ito:2016aai,Ito:2017bnn,Do:2017qyd,Do:2017onf,Do:2018zac,Franciolini:2018eno,Gong:2019hwj,Do:2021lyf,Gorji:2020vnh,Firouzjahi:2018wlp} in order to explore the early universe.
The original anisotropic inflation model~\cite{Watanabe:2009ct} is endowed with a U(1) gauge field.
In high-energy fundamental theories, we can expect the existence of multiple U(1) gauge fields in the early universe~\cite{Yamamoto:2012tq}.
Interestingly, it was shown that multiple U(1) gauge fields tend to select a minimally anisotropic configuration dynamically~\cite{Yamamoto:2012tq}.
A two-form field, which is also a gauge field, can give rise to a prolate-type anisotropy opposed to an oblate-type anisotropy from a U(1) gauge field~\cite{Ohashi:2013mka,Ohashi:2013qba,Ito:2015sxj}.
Moreover, an SU(2) gauge field coupled to an inflaton in the axially symmetric Bianchi type I spacetime was studied in \cite{Murata:2011wv}.
It was shown that an SU(2) gauge field could result in both prolate- and oblate-type anisotropies.
In general, nonlinear self-couplings in the kinetic term of an SU(2) gauge field cause the decay of the SU(2) gauge field after sufficient growth~\cite{Maeda:2012eg,Maeda:2013daa}.
That behavior would enrich the predictions for observations and 
support the cosmic no-hair conjecture.

In the standard model of particle physics, not only U(1) and SU(2) gauge fields but also an SU(3) gauge field plays an important role. In high-energy fundamental theories, there are many
non-Abelian gauge fields including SU(3) and other Lie groups.
Therefore, it would be interesting to study the role of a non-Abelian gauge field in the early universe in addition to the previous works for the cases of U(1) or SU(2) gauge fields~\cite{Yamamoto:2012tq,Murata:2011wv,Maeda:2012eg,Maeda:2013daa}.
In this paper, as a first step, we study inflationary universes with an SU(3) gauge field, which we do not identify as the one in the standard model, coupled to an inflaton in the Bianchi type I spacetime.
As we will see, nonlinear self-couplings in the kinetic term of the SU(3) gauge field make inflationary dynamics complicated as well as the case of an SU(2) gauge field~\cite{Maeda:2012eg,Maeda:2013daa}.
In generic setups, namely, when all the initial values for the components of the SU(3) gauge field have the same order of magnitude, all the components eventually decay due to the nonlinear self-couplings.
This result supports the cosmic no-hair conjecture.
However, the transient anisotropy can be important observationally. 
Moreover, we find  specific cases where two components of the SU(3) gauge field survive even in the presence of the nonlinear self-couplings and the anisotropic expansion of the inflationary universe lasts.
This is quite in contrast to the result in the case of an SU(2) gauge field~\cite{Maeda:2012eg,Maeda:2013daa} 
and the behavior is also different from that of two U(1) gauge fields~\cite{Yamamoto:2012tq}.
The reason can be attributed to the flat direction of potential of gauge fields. We argue that this effect occurs for general non-Abelian gauge fields except for an SU(2) gauge field.

The organization of the paper is as follows.  
In \S\ref{sec:U1}, we review the anisotropic power-law inflation with U(1) gauge field(s).
In \S\ref{sec:SU3}, we study the anisotropic power-law inflation with an SU(3) gauge field.  
We first study the case of SU(2)\:$\otimes$\:U(1). 
Secondly, we investigate a specific example with a particular initial condition of the SU(3) gauge field. 
It turns out that the results are different from those of an SU(2) gauge field or two U(1) gauge fields. 
At last, we study the case where all the initial values for the components of the SU(3) gauge field have the same order of magnitude.
In \S\ref{sec:Extension}, we discuss inflation with general non-Abelian gauge fields.
The final section~\ref{sec:conclusion} is devoted to conclusion.

\section{Inflation with U(1) Gauge Field(s)}  \label{sec:U1}

In this section, we briefly review three scenarios of inflation, where U(1) gauge fields are coupled to an inflaton~$\phi$ through an exponential-type gauge kinetic function.
In each case, there exists an exact solution, which would be useful when we discuss the attractor behavior in the case of an SU(3) gauge field later.

\subsection{Anisotropic power-law inflation with a U(1) gauge field}
\label{ssec_U(1)}

Let us first consider the model with a single U(1) gauge field, which can be regarded as a subgroup of SU(3).
In this case, the action is given by
\begin{equation}
S=\int \D^4x  \sqrt{-g} \brb{ \frac{\Mpl^2} {2} R -\frac{1}{2} \na_\mu\phi\na^\mu\phi -V(\phi)  -\frac{1}{4} f^2(\phi) F_{\mu\nu} F^{\mu\nu}} ,
\end{equation}
where $g$ is the determinant of the metric, $R$ is the Ricci scalar, $F_{\mu\nu}\coloneqq \pa_\mu A_\nu-\pa_\nu A_\mu$ is the field strength of the U(1) gauge field $A_\mu$, and $\Mpl$ denotes the reduced Planck mass.
We assume the potential~$V(\phi)$ and the gauge kinetic function~$f(\phi)$ respectively have the form
    \be
    V(\phi)=V_0\exp\bra{\la\fr{\phi}{\Mpl}}, \qquad
    f(\phi)=f_0\exp\bra{\rho\fr{\phi}{\Mpl}}, \label{Vandf}
    \ee
with $V_0$, $f_0$, $\la$, and $\rho$ being positive constants.
We introduce dimensionless quantities as
    \be
    \hat{x}^\mu\coloneqq \Mpl x^\mu, \qquad
    \hat{V}_0\coloneqq \fr{V_0}{\Mpl^4}, \qquad
    \hat{\phi}\coloneqq \fr{\phi}{\Mpl}, \qquad
    \hat{A}_{\mu} \coloneqq \fr{A_{\mu}}{\Mpl}. \label{nondim}
    \ee
Then, the model is characterized by the four parameters~$\hat{V}_0$, $f_0$, $\lambda$, and $\rho$.
In what follows, we omit hats from the dimensionless quantities for notational convenience.

The authors of \cite{Kanno:2010nr} studied an exact solution of anisotropic power-law inflation in this model. 
They assumed a homogeneous spacetime and fields of the form,
    \be
    \begin{split}
    &\D s^2=-\D t^2+\E^{2\alpha(t)}\brb{\E^{2\beta(t)}(\D x^2+\D y^2)+\E^{-4\beta(t)}\D z^2}, \qquad
    \phi=\phi(t), \qquad
    A_\mu \D x^\mu=A_3(t)\D z,
    \end{split}
    \ee
and showed that the following configuration solves the system of equations of motion~(EOMs):
    \be
    \begin{split}
    \alpha=\zeta\ln t, \qquad
    \beta=\eta\ln t, \qquad
    \phi=-\fr{2}{\la}\ln t+\phi_0, \qquad
    \dot{A}_3=f^{-2}(\phi)\E^{-\alpha-4\beta}p_A.
    \end{split}
    \ee
Here, we have defined
    \be
    \begin{split}
    \zeta\coloneqq \fr{\la^2+8\la\rho+12\rho^2+8}{6\la(\la+2\rho)}, \qquad
    \eta\coloneqq \fr{\la^2+2\la\rho-4}{3\la(\la+2\rho)},
    \end{split}
    \ee
and the values of $\phi_0$ and $p_A$ are determined from
    \be
    \begin{split}
    V_{0} \E^{\lambda \phi_{0}} &=\frac{\left(\lambda\rho+2 \rho^{2}+2\right)\left(-\lambda^{2}+4 \lambda\rho+12 \rho^{2}+8\right)}{2 \lambda^{2}(\lambda+2 \rho)^{2}}
    \eqqcolon u, \\
    p_{A}^{2} f_{0}^{-2} \E^{-2 \rho \phi_{0}} &=\frac{\left(\lambda^{2}+2 \lambda\rho-4\right)\left(-\lambda^{2}+4 \lambda\rho+12 \rho^{2}+8\right)}{2 \lambda^{2}(\lambda+2 \rho)^{2}}
    \eqqcolon w.
    \end{split}
    \ee
Note that $V_0\E^{\la\phi_0}$ and $p_A^2f_0^{-2}\E^{-2\rho\phi_0}$ are intrinsically positive, and hence this type of solution exists only if $u>0$ and $w>0$.

For this solution, we obtain the Hubble parameter and the slow-roll parameter as
    \be 
    H\coloneqq\dot{\alpha}=\fr{\zeta}{t}, \qquad
    \ep\coloneqq -\fr{\dot{H}}{H^2}
    =\fr{1}{\zeta}=\fr{6\la(\la+2\rho)}{\la^2+8\la\rho+12\rho^2+8}, \label{eq:epsilon}
    \ee
where a dot denotes a derivative with respect to $t$.
Hence, if we choose $\la$ sufficiently small, then we have $\ep\ll 1$, i.e., an inflationary universe can be realized.
For a small enough $\la$, the condition~$u>0$ is trivially satisfied, while we need $\la^2+2\la\rho>4$ to guarantee $w>0$.
Also, the following parameter is useful to measure the anisotropy:
\be  \label{sigma_U1}
\sigma \coloneqq \frac{\dot{\beta}}{H}=\fr{\eta}{\zeta}
=\frac{2\left(\lambda^{2}+2 \lambda\rho -4\right)}{\lambda^{2}+8 \lambda
\rho+12 \rho^{2}+8}.
\ee
Finally, for later reference, let us compute the density parameter for the gauge field.
The energy density of the gauge field is given by
    \be
    \rho_{g}
    =\fr{1}{2}f^2g^{33}(\dot{A}_3)^2,
    \ee
and hence the density parameter is written as
\be 
\Omega_{g} 
\coloneqq \fr{\rho_{g}}{3H^2} = \fr{w}{6\zeta^2}
=\fr{3  \left(\lambda^{2}+2 \lambda\rho -4\right)\left(-\lambda^{2}+4 \lambda\rho +12 \rho^{2}+8\right)}{(\lambda^{2}+8 \lambda\rho +12 \rho^{2}+8 )^2}.  \label{Omega_U1}
\ee

\subsection{Anisotropic power-law inflation with two U(1) gauge fields}
\label{ssec_2U(1)}

Next, let us consider the model with two copies of U(1) gauge fields~$A^{(1)}_\mu$ and $A^{(2)}_\mu$, which can also be embedded into SU(3).
The action is written as
\begin{equation}
S=\int \D^4x \sqrt{-g} \brb{ \frac{\Mpl^2} {2} R -\frac{1}{2} \na_\mu\phi\na^\mu\phi -V(\phi)  -\frac{1}{4} f^2(\phi) \sum_{n=1}^2 F^{(n)}_{\mu\nu} F^{(n)\mu\nu}},
\end{equation}
with $F^{(n)}_{\mu\nu}\coloneqq \pa_\mu A^{(n)}_\nu-\pa_\nu A^{(n)}_\mu$ being the field strength of the $n$th gauge field.
Also, the potential~$V(\phi)$ and the gauge kinetic function~$f(\phi)$ are of the same form as in \eqref{Vandf}.

From \cite{Yamamoto:2012tq}, we know there exists a stable fixed point with an orthogonal configuration
of the two U(1) gauge fields, where the spacetime and fields are of the form, 
    \be
    \begin{split}  
    &\D s^2 =-\D t^2+\E^{2\alpha(t)}\brb{\E^{2\beta(t)}(\D x^2+\D y^2)+\E^{-4\beta(t)}\D z^2}, \\
    &\phi=\phi(t), \qquad  
    A^{(1)}_\mu \D x^\mu  =A^{(1)}_1(t)\D x, \qquad
    A^{(2)}_\mu \D x^\mu=A^{(2)}_2(t)\D y, 
    \end{split}
    \ee
where
    \be
    \alpha=\fr{  (\la + 2\rho)(\la+6\rho)+2}   {3\la(\la+4\rho)} \ln t, \qquad
    \beta=-\fr{\la^2+2\la\rho-4}{3\la(\la+4\rho)} \ln t, \qquad
    \phi=-\fr{2}{\la}\ln t + \phi_0 ,
    \ee
with $\phi_0$ being constant.
The anisotropy is given by 
\be
\sigma = \frac{\dot{\beta}}{H}= - \fr{\la^2+2\la\rho-4}{(\la + 2\rho)(\la+6\rho)+2}, 
\ee
and the total energy density of the gauge fields is given by
    \be
    \rho_{g} = \fr{1}{2}\sum_{n=1}^2  f^2g^{ij}\dot{A}^{(n)}_i\dot{A}^{(n)}_j
    =\fr{18(\la^2+2\la\rho-4)(6\rho^2+2\la\rho+1)}{[(\la + 2\rho)(\la+6\rho)+2]^2} H^2.
    \ee
Here, $H=\dot{\alpha}$ is the Hubble parameter and each gauge field shares half of the total energy density.
Hence, the density parameter for the gauge fields is written as
    \be
    \Omega_{g}=\fr{\rho_g}{3H^2}
    =\fr{6(\la^2+2\la\rho-4)(6\rho^2+2\la\rho+1)}{[(\la + 2\rho)(\la+6\rho)+2]^2}. 
    \ee

\subsection{Isotropic power-law inflation with multiple U(1) gauge fields}
\label{ssec_multiU(1)}

Finally, let us consider the model with $N$ copies of U(1) gauge fields~$A^{(n)}_\mu$ ($n=1,\cdots,N$), with $N\ge 3$.
These multiple U(1) cases cannot be embedded into SU(3).
Nevertheless, it is useful to compare the generic behavior of the model with the SU(3) gauge field model.
Now, the action is written as
\begin{equation}
S=\int \D^4x \sqrt{-g} \brb{ \frac{\Mpl^2} {2} R -\frac{1}{2} \na_\mu\phi\na^\mu\phi -V(\phi)  -\frac{1}{4} f^2(\phi) \sum_{n=1}^N F^{(n)}_{\mu\nu} F^{(n)\mu\nu}}, \label{action_multiU(1)}
\end{equation}
with $F^{(n)}_{\mu\nu}\coloneqq \pa_\mu A^{(n)}_\nu-\pa_\nu A^{(n)}_\mu$ being the field strength of the $n$th gauge field.
Also, the potential~$V(\phi)$ and the gauge kinetic function~$f(\phi)$ are of the same form as in \eqref{Vandf}.
The authors of \cite{Yamamoto:2012tq} studied a more general case where each gauge field is coupled to $\phi$ with a different coupling constant (i.e., different $\rho$ for different $n$).
However, we restrict ourselves to the model described by the action~\eqref{action_multiU(1)} for simplicity.
Note also that we use dimensionless quantities similar to those in \eqref{nondim} in the following discussion.

As we did in the previous section, one can study power-law inflation in the present model.
It was shown in \cite{Yamamoto:2012tq} that there exists isotropic stable fixed points with nontrivial configuration of the gauge fields, where the spacetime and fields are of the form,
    \be
    \begin{split}
    &\D s^2=-\D t^2+\E^{2\alpha(t)}\bra{\D x^2+\D y^2+\D z^2}, \qquad
    \phi=\phi(t), \qquad
    A^{(n)}_\mu \D x^\mu=A^{(n)}_i(t)\D x^i.
    \end{split}
    \ee
where
    \be
    \alpha=\fr{\la+2\rho}{2\la}\ln t, \qquad
    \phi=-\fr{2}{\la}\ln t+\phi_0, 
    \ee
with $\phi_0$ being constant, and the total energy density of the gauge fields is given by
    \be
    \rho_{g} = \fr{1}{2}\sum_{n=1}^N  f^2g^{ij}\dot{A}^{(n)}_i\dot{A}^{(n)}_j
    =\fr{3(\la^2+2\la\rho-4)}{(\la+2\rho)^2}H^2.
    \ee
 Here, $H=\dot{\alpha}$ is the Hubble parameter.
Hence, the density parameter for the gauge fields is written as
    \be
    \Omega_{g}=\fr{\rho_g}{3H^2}
    =\fr{\la^2+2\la\rho-4}{(\la+2\rho)^2}. \label{Omega_multi-U1}
    \ee
Note that this value is the same for any $N\ge3$. 
In the case of $N=3$, each gauge field shares one-third of the total energy density.

\section{Inflation with an SU(3) Gauge Field}  \label{sec:SU3}

In this section, we study an inflationary universe with an SU(3) gauge field~$A^a_\mu$.
The SU(3) gauge field is written in the form
\begin{align}
\mathbf{A}=A^a_{\mu} T^a \D x^\mu ,
\end{align}
where $T^a$'s are the SU(3) generators defined by $T^a = \lambda^a/2$ with the Gell-Mann matrices~$\la^a$:
\be
\begin{split}
&
\lambda^1 
  =
\begin{pmatrix}
~0 & 1 & 0 ~\\
~1 & 0 & 0~\\
~0 & 0 & 0  ~\\
\end{pmatrix}
,\quad 
\lambda^2 
  =
\begin{pmatrix}
~0 & -i & 0 ~\\
~i & 0 & 0~\\
~0 & 0 & 0  ~\\
\end{pmatrix}
,\quad 
\lambda^3 
  =
\begin{pmatrix}
~1 & 0 & 0 ~\\
~0 & -1 & 0~\\
~0 & 0 & 0  ~\\
\end{pmatrix}
,\quad 
\lambda^4 
  =
\begin{pmatrix}
~0 & 0 & 1 ~\\
~0 & 0 & 0~\\
~1 & 0 & 0  ~\\
\end{pmatrix},
\\
& \lambda^5 
  =
\begin{pmatrix}
~0 & 0 & -i ~\\
~0 & 0 & 0~\\
~i & 0 & 0  ~\\
\end{pmatrix}
, \quad 
\lambda^6 
  =
\begin{pmatrix}
~0 & 0 & 0 ~\\
~0 & 0 & 1~\\
~0 & 1 & 0  ~\\
\end{pmatrix}
,\quad
\lambda^7 
  =
\begin{pmatrix}
~0 & 0 & 0 ~\\
~0 & 0 & -i~\\
~0 & i & 0  ~\\
\end{pmatrix}
,\quad 
\lambda^8 
  =
\frac{1}{\sqrt{3}}
\begin{pmatrix}
~1 & 0 & 0 ~\\
~0 & 1 & 0~\\
~0 & 0 & -2  ~\\
\end{pmatrix}
.
\end{split}
\ee
The generator matrices satisfy the normalization condition
 \begin{align}
 \mathrm{Tr}\,(T^a T^b) =\fr{1}{2} \delta^{ab},
 \end{align}
and the commutation relation 
\begin{align}
\left[T^a, T^b \right]=  i f^{a b c} T^c.
\end{align} 
Here, $f^{abc}$ is the structure constant satisfying
\begin{align}
f^{a b c}= - 2 i\,\mathrm{Tr}\left( T^a \left[T^b, T^c \right]\right),
\end{align}
which is completely antisymmetric.
The nonvanishing components of $f^{abc}$ are 
\begin{align}
f^{123}=1, \qquad 
f^{147}=f^{165}=f^{246}=f^{257} = f^{345}=f^{376}={\frac {1}{2}}, \qquad 
f^{845}=f^{867}={\frac {\sqrt {3}}{2}}.
\end{align}
The field strength of the gauge field is given by
\begin{align}
F^a_{\mu\nu}  = \nabla_{\mu} A^a_{\nu} -\nabla_{\nu} A^a_{\mu} + g_*  f^{abc} A^b_{\mu} A^c_{\nu}, \label{field_strength}
\end{align}
where $g_*$ is the gauge coupling constant. 

Now, we are ready to write down the action.
Similarly to the model with a U(1) gauge field studied in \cite{Kanno:2010nr}, we study a model with an SU(3) gauge field described by the following action:
\begin{equation}\label{action}
S=\int \D ^4x  \sqrt{-g} \brb{ \frac{\Mpl^2} {2} R -\frac{1}{2} \na_\mu\phi\na^\mu\phi -V(\phi) -\frac{1}{4} f^2(\phi) F^a_{\mu\nu} F^{a\mu\nu}}.
\end{equation}
Here, the potential~$V(\phi)$ and the gauge kinetic function~$f(\phi)$ are of the same form as in \eqref{Vandf}, which we reproduce here for the convenience:
    \be
    V(\phi)=V_0\exp\bra{\la\fr{\phi}{\Mpl}}, \qquad
    f(\phi)=f_0\exp\bra{\rho\fr{\phi}{\Mpl}},
    \ee
with $V_0$, $f_0$, $\la$, and $\rho$ being positive constants.
As we did in \S\ref{ssec_U(1)}, we introduce dimensionless quantities as follows:
    \be
    \hat{x}^\mu\coloneqq \Mpl x^\mu, \qquad
    \hat{V}_0\coloneqq \fr{V_0}{\Mpl^4}, \qquad
    \hat{f}_0\coloneqq \fr{f_0}{g_*}, \qquad
    \hat{\phi}\coloneqq \fr{\phi}{\Mpl}, \qquad
    \hat{A}^a_{\mu} \coloneqq \fr{g_*}{\Mpl}A^a_{\mu}. \label{nondim_SU(3)}
    \ee
Note that the gauge coupling constant~$g_*$ has been absorbed into the field redefinition.
Moreover, one can set $\hat{f}_0\to 1$ by shifting $\hat{\phi}\to \hat{\phi}-\rho^{-1}\ln \hat{f}_0$.
Then, the model is characterized by the three parameters~$\hat{V}_0$, $\lambda$, and $\rho$.
In what follows, we suppress hats for notational convenience.

It is straightforward to obtain the EOMs. 
The Einstein equations read
\be
R_{\mu\nu} = T_{\mu\nu}  - \dfrac{1}{2} g_{\mu\nu} T^\rho{}_\rho, 
\ee
where $R_{\mu\nu}$ is the Ricci tensor and 
$T_{\mu\nu}= T^{\phi}_{\mu\nu} + T^g_{\mu\nu}$ is the energy-momentum tensor.
Here, $T^{\phi}_{\mu\nu}$ and $T^{g}_{\mu\nu}$ denote the contribution
from the scalar 
\be
T^{\phi}_{\mu\nu} = \nabla_{\mu}\phi \nabla_{\nu} \phi -\dfrac{1}{2} g_{\mu\nu} \nabla_{\rho}\phi \nabla^{\rho}\phi - g_{\mu\nu} V, 
\ee
and that from the gauge field
\be
T^g_{\mu\nu} = f^2(\phi) \bra{-\dfrac{1}{4} g_{\mu \nu} F^a_{\rho\sigma}F^{a \rho\sigma} + F^a_{\mu \sigma} F_{\nu}^{a \sigma}} .
\ee
The EOM for the scalar is given by
\be
-\nabla_{\mu}\nabla^{\mu}  \phi +\dfrac{\D V}{\D \phi} = -\dfrac{1}{4} F^a_{\mu\nu}F^{a \mu\nu} \dfrac{\D f^2(\phi)}{\D \phi} .
\ee
The EOMs for the gauge fields are
\be
\nabla_{\mu} F^{a\mu\nu} + f^{abc} A^b_{\mu} F^{c\mu\nu} = - \dfrac{\nabla_{\mu} f^2(\phi)}{f^2(\phi)} F^{a\mu\nu} . \label{EOMs_gauge}
\ee

We study a general Bianchi Type I universe having the metric of the form
    \be
    g_{\mu\nu} \D x^{\mu}\D x^{\nu}=-\D t^2+ g_{ij}(t) \D x^i \D x^j,
    \ee
accompanied by homogeneous scalar and gauge fields,
    \be
    \phi=\phi(t), \qquad
    A^a_\mu \D x^\mu=A^a_i(t)\D x^i,
    \ee
where we used the gauge symmetry to fix the time component of the gauge field.
In this setup, the system of EOMs consists of the Hamiltonian constraint, 
6 EOMs for $g_{ij}$, 8 Yang-Mills constraints, 24 EOMs for  $A^a_i$, 1 EOM for $\phi$.
Namely, 9 constraint equations and 31 second-order ordinary differential equations in total.
The Hamiltonian and the Yang-Mills constraints are used to provide a consistent set of initial data.

Let us introduce some useful notations.
We define the Hubble parameter~$H$ by 
\begin{align}
H \coloneqq \dot{\alpha}, \qquad
\alpha \coloneqq \dfrac{1}{6} \mathrm{ln}(\mathrm{det}(g_{ij})). \label{eq:alpha}
\end{align}
The e-folding number~$N$ is given by the change in the parameter~$\alpha$.
Also, we define a matrix $(\E^{2\beta})_{ij}$  by
\begin{align}
\bra{\E^{2\beta}}_{ij} \coloneqq \E^{-2\alpha}g_{ij}.
\end{align}
From \eqref{eq:alpha}, we know that the determinant of $(\E^{2\beta})_{ij}$ is unity, or equivalently, $\beta_{ij}$ is traceless.
In terms of this $(\E^{2\beta})_{ij}$, we define the anisotropy matrix by 
\begin{align}
\sigma_{ij}\coloneqq \fr{1}{2H}\bra{\E^{-2\beta}\fr{\D \E^{2\beta}}{\D t}}_{(ij)},
\end{align}
where we have denoted the symmetrization of two indices by $(ij)$. 
Also, we introduce the root-mean-square anisotropy as 
\begin{align}
\si\coloneqq \sqrt{\fr{1}{6}\sum_{i,j=1}^{3}\si_{ij}\si_{ij}}\ .
\end{align}
Note that, if $\sigma=0$, then all the components of $\sigma_{ij}$ must vanish, which is nothing but the isotropic case.

In order to study the dynamics of the spacetime and the gauge field, it is useful to define the density parameters for the gauge field.
In the present setup, the field strength takes the form
    \be
    F^a_{0j}=\dot{A}^a_j, \qquad
    F^a_{ij}=f^{abc}A^b_iA^c_j,
    \ee
which we call the electric part and the magnetic part, respectively.
Then, the energy density of the gauge field is written as
    \be
    \rho_g\coloneqq T^g_{00}
    =f^2(\phi)\bra{\fr{1}{2}g^{ij}\dot{A}^a_i\dot{A}^a_j+\fr{1}{4}f^{abc}f^{ade}A^b_iA^c_jA^{di}A^{ej}},
    \ee
which can be separated into the contributions from the electric and the magnetic parts, i.e.,
    \be
    \rho_E\coloneqq \fr{f^2(\phi)}{2}g^{ij}\dot{A}^a_i\dot{A}^a_j, \qquad
    \rho_B\coloneqq f^2(\phi)V_g. \label{em}
    \ee
Here, $V_g$ is defined by
\begin{eqnarray}
V_g&\coloneqq& \fr{1}{4}f^{abc}f^{ade}A^b_iA^c_jA^{di}A^{ej} \nonumber \\
&=& \left(A^{2}_{[i} A^{3}_{j]} + \frac{1}{2}A^{4}_{[i} A^{7}_{j]}
+ \frac{1}{2}A^{6}_{[i} A^{5}_{j]} \right)^2
+\left(A^{3}_{[i} A^{1}_{j]} + \frac{1}{2}A^{4}_{[i} A^{6}_{j]}
+ \frac{1}{2}A^{5}_{[i} A^{7}_{j]} \right)^2
+\left(A^{1}_{[i} A^{2}_{j]} + \frac{1}{2}A^{4}_{[i} A^{5}_{j]}
+ \frac{1}{2}A^{7}_{[i} A^{6}_{j]} \right)^2   \nonumber\\
&& +\frac{1}{4}\left[A^{1}_{[i} A^{7}_{j]} + A^{2}_{[i} A^{6}_{j]}
  + \left( A^{3}_{[i}  + \sqrt{3} A^{8}_{[i} \right) A^{5}_{j]}\right]^2
+\frac{1}{4}\left[A^{1}_{[i} A^{6}_{j]} + A^{7}_{[i} A^{2}_{j]}
+ \left( A^{3}_{[i}  + \sqrt{3} A^{8}_{[i} \right) A^{4}_{j]}\right]^2 \nonumber\\
&& +\frac{1}{4}\left[A^{5}_{[i} A^{1}_{j]} + A^{2}_{[i} A^{4}_{j]}
+ \left( A^{3}_{[i} - \sqrt{3}  A^{8}_{[i} \right) A^{7}_{j]} \right]^2
+\frac{1}{4}\left[A^{1}_{[i} A^{4}_{j]} + A^{2}_{[i} A^{5}_{j]}
- \left( A^{3}_{[i} - \sqrt{3} A^{8}_{[i} \right) A^{6}_{j]}\right]^2 \nonumber\\
&& +\frac{3}{4}\left(A^{4}_{[i} A^{5}_{j]} + A^{6}_{[i} A^{7}_{j]}  \right)^2 , \label{V}
\end{eqnarray}
which amounts to the potential for the gauge field.
Note that square brackets~$[ij]$ denote antisymmetrization and we have denoted $(B_{ij})^2\coloneqq B_{ij}B^{ij}$ for an arbitrary quantity~$B_{ij}$ with spatial indices.
Now, we can define the density parameter for each component, 
    \be
    \Omega_E\coloneqq \fr{\rho_E}{3H^2}, \qquad
    \Omega_B\coloneqq \fr{\rho_B}{3H^2},
    \ee
so that the total density parameter for the gauge field is given by $\Omega_T\coloneqq \Omega_E+\Omega_B$. 
It should be noted that $\Omega_B$ measures the effect of nonlinear self-interactions.
It is also useful to define the electric-part density parameter for each $a$, i.e.,
    \be
    \Omega^a\coloneqq \fr{f^2(\phi)}{6H^2}g^{ij}\dot{A}^a_i\dot{A}^a_j\qquad \text{(no sum over $a$)},
    \ee
so that $\sum_{a}\Omega^a=\Omega_E$.

We are now ready to investigate the evolution of inflationary universes with an SU(3) gauge field.
It is useful to start with the simplest case and go step by step.
The simplest subgroup of SU(3) is U(1). There are also U(1)\:$\otimes$\:U(1) and SU(2) subgroups. These cases have been already studied.
We shall start with the next simplest case~SU(2)\:$\otimes$\:U(1) and
proceed step by step to explore the inflationary universe with an SU(3) gauge field.
In numerical computations, we put the initial time to be $t=1$ and set $A^a_{i}=0$ so that the Yang-Mills constraints are satisfied.
For a given set of parameters~$(V_0,\la,\rho)$, we fix initial values for $\phi$, $\dot{\phi}$, and the velocity of the gauge field by use of the exact solutions mentioned in the previous section.
We take $\la=0.8$ and $\rho=4$, for which the anisotropic inflation is realized in the U(1) model~(see \S\ref{ssec_U(1)}).
In \S\ref{sec_su2*u1} and \S\ref{sec_specific}, as for the spatial part of the metric, we assume $g_{ij}=\delta_{ij}$ and $\dot{g}_{ij}=2H_{\rm in}\delta_{ij}$ at the initial time, where the value of the constant~$H_{\rm in}$ is determined from the Hamiltonian constraint.

\subsection{\texorpdfstring{SU(2)\:$\otimes$\:U(1) subgroup}{SU(2) times U(1) subgroup}}  \label{sec_su2*u1}

In this subsection, we investigate SU(2)\:$\otimes$\:U(1) subgroup of 
the SU(3) gauge field.
Namely, we consider the case where only $A^1_{1}$, $A^2_{2}$, $A^3_{3}$, and $A^8_{3}$ are nonvanishing.
Moreover, we impose the axial symmetry along the $z$-direction, so that the spacetime and the gauge field have the following form:
\be
    \D s^2=  -\D t^2 +  g_{11} \left(\D x^2 +  \D y^2\right)  +g_{33} \D z^2 , \qquad
    A^1_{1}=A^2_{2}.
\ee
In fact, as is shown in the \hyperref[Appendix]{Appendix}, we can classify the gauge-field configurations which are consistent with the axial symmetry.
Those classes of configurations can be treated in a similar manner.

Now, the number of EOMs reduces to seven.
All the other EOMs become trivial. 
Performing the transformation of variables  
\be
g_{11} = \exp(2\alpha+2\beta), \qquad
g_{33} = \exp(2\alpha-4\beta),
\ee
we obtain the Hamiltonian constraint
\be
  3  \left( -\dot{\alpha}^{2}  +\dot{\beta}^{2} \right)  +\frac{1}{2} \dot{\phi}^{2}+V  
+\frac{1}{2} \E^{-2\alpha}f^{2}\left[
 2 \E^{-2 \beta} \dot{A}_{11}^{2}
+   \E^{4 \beta} (\dot{A}_{33}^{2} +\dot{A}_{83}^{2} )
+2 
\E^{-2 \alpha+2 \beta} A_{11}^{2}   A_{33}^{2}
+
\E^{-2 \alpha-4 \beta} A_{11}^{4}  \right] =0   ,  
\ee
the Einstein equations 
\be
\begin{split}
 2 \ddot{\alpha} +3\left(\dot{\alpha}^{2}+\dot{\beta}^{2}\right)+\frac{1}{2} \dot{\phi}^{2}-V 
 +\frac{1}{6} \E^{-2\alpha}f^{2}\left[
  2 \E^{-2 \beta} \dot{A}_{11}^{2}
  + \E^{4 \beta} (\dot{A}_{33}^{2} +\dot{A}_{83}^{2} )
 +2 \E^{-2 \alpha+2 \beta} A_{11}^{2}   A_{33}^{2}
 + \E^{-2 \alpha-4 \beta} A_{11}^{4}
 \right]&=0   , \\
\ddot{\beta} +3 \dot{\alpha} \dot{\beta} 
 -\frac{1}{3} \E^{-2\alpha}f^{2}\left[
 \E^{4 \beta} (\dot{A}_{33}^{2} +\dot{A}_{83}^{2} )
 -\E^{-2 \beta} \dot{A}_{11}^{2}
 - \E^{-2 \alpha+2 \beta} A_{11}^{2} A_{33}^{2}
 + \E^{-2 \alpha-4 \beta} A_{11}^{4}
 \right]&=0  , 
\end{split}
\ee
the EOM for the inflaton
\be
\ddot{\phi} +3 \dot{\alpha} \dot{\phi}+V^{\prime} 
 -\E^{-2\alpha}f f^{\prime}\left[
 \E^{4 \beta} (\dot{A}_{33}^{2} +\dot{A}_{83}^{2} )
 +2 \E^{-2 \beta} \dot{A}_{11}^{2}
 -2 \E^{-2 \alpha+2 \beta} A_{11}^{2} A_{33}^{2}
 - \E^{-2 \alpha-4 \beta} A_{11}^{4}
 \right] =0  ,
\ee
and the EOMs for the gauge field
\be
\begin{split}
 \ddot{A}_{11}+2 \frac{f^{\prime}}{f} \dot{\phi} \dot{A}_{11}+(\dot{\alpha}-2 \dot{\beta}) \dot{A}_{11}+
 \E^{-2 \alpha+4 \beta} A_{33}^{2} A_{11}+
 \E^{-2 \alpha-2 \beta} A_{11}^{3}&=0 , \\
\ddot{A}_{33}+2 \frac{f^{\prime}}{f} \dot{\phi} \dot{A}_{33}+(\dot{\alpha}+4 \dot{\beta}) \dot{A}_{33}+2
\E^{-2 \alpha-2 \beta} A_{11}^{2} A_{33}&=0 ,\\
\ddot{A}_{83}+2 \frac{f^{\prime}}{f} \dot{\phi} \dot{A}_{83}+(\dot{\alpha}+4 \dot{\beta}) \dot{A}_{83}&=0 , 
\end{split}
\ee
where 
$f'\coloneqq \D f/\D\phi$ and we lowered the gauge index~$a$ for gauge-field components~$A^a_i$ for notational convenience.
As the component~$A^8_{3}$ is decoupled from the SU(2) sector, its EOM can be immediately integrated to yield
\begin{align}
\dot{A}_{83} =  f^{-2} \E^{-\alpha-4\beta} p_{83},
\end{align}
where $p_{83}$ is a constant.

\begin{figure}[H]
\includegraphics[clip, width=0.5\columnwidth]{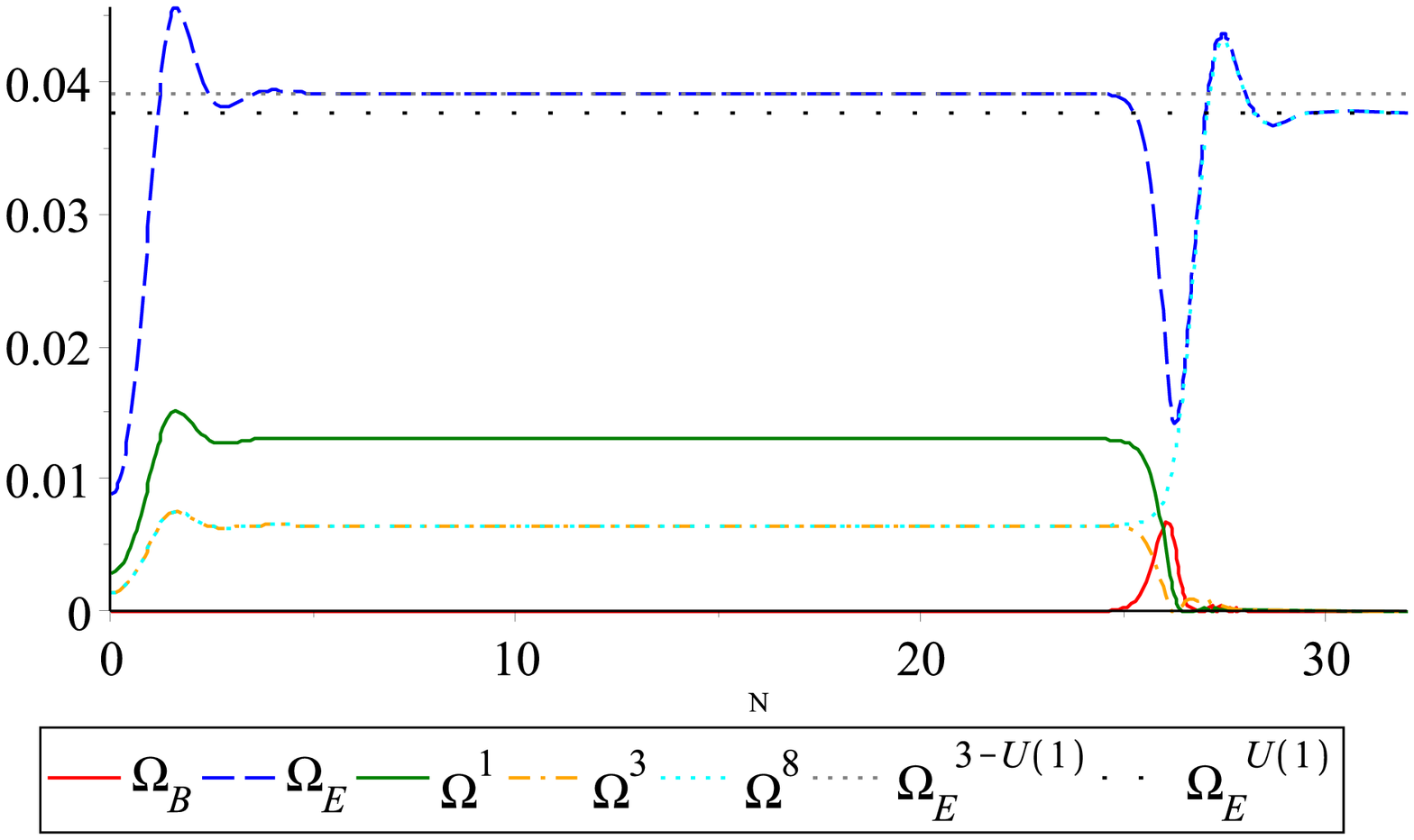} 
\includegraphics[clip, width=0.3\columnwidth]{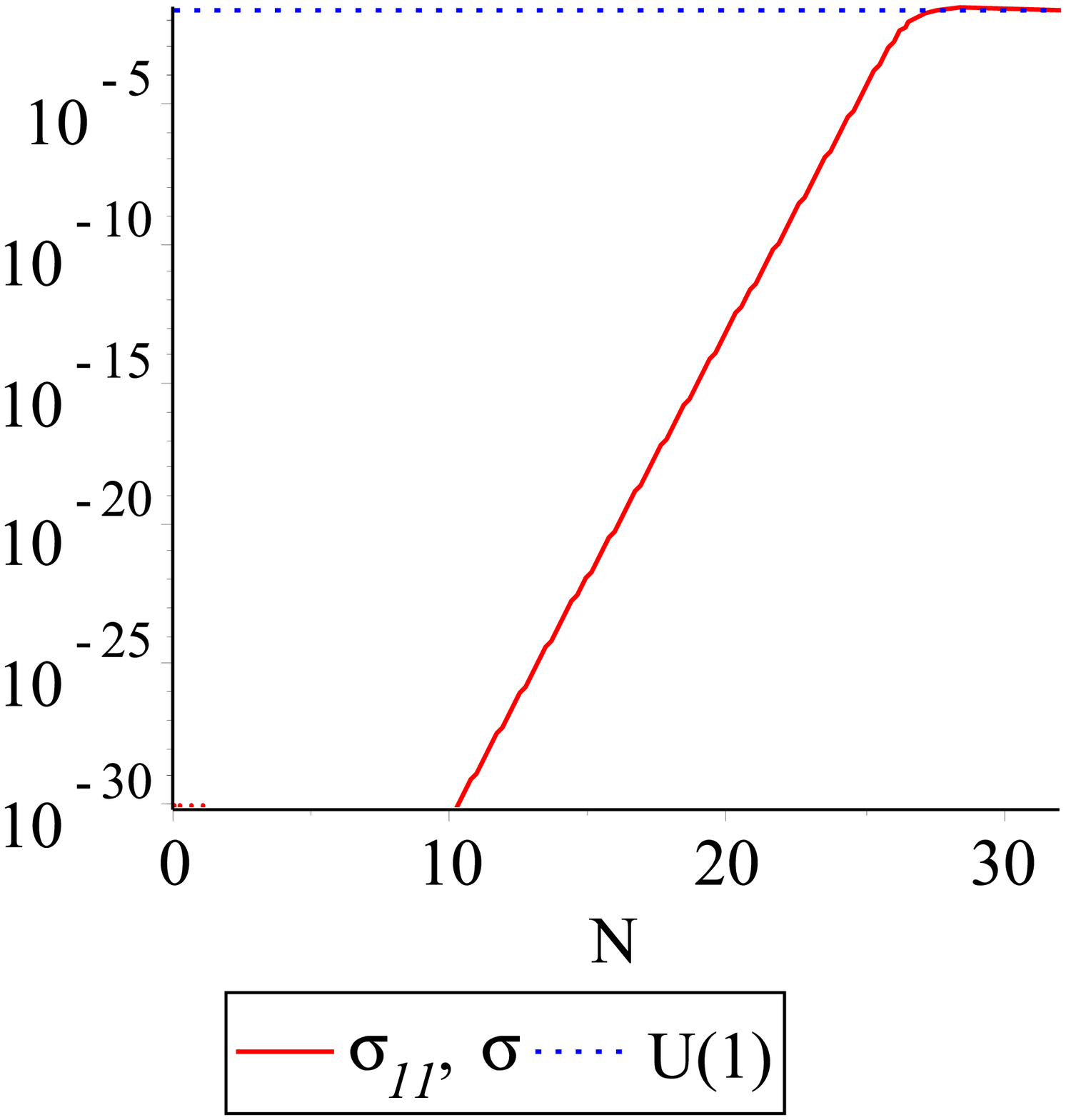} 
\caption{Evolution of the density parameters of gauge-field components (left) and anisotropy (right) against the number of e-folds for an initial condition with $\dot{A}^1_{1}=\dot{A}^2_{2}=\sqrt{2}\dot{A}^3_{3}=\sqrt{2}\dot{A}^8_{3}$. 
In the left graph, the red solid, blue dashed, green solid, orange dash-dotted, and cyan dotted curves respectively correspond to $\Omega_B$, $\Omega_E$, $\Omega^{1}$, $\Omega^{3}$, and $\Omega^{8}$.
The gray dotted and black space-dotted lines represent $\Omega_E$ for the isotropic three-U(1) case~\eqref{Omega_multi-U1} and $\Omega_E$ for the one-U(1) case~\eqref{Omega_U1}, respectively. 
In the right graph, the red solid curve corresponds to $\sigma_{11}(=\sigma)$ and the blue dotted line represents the anisotropy for the one-U(1) case~\eqref{sigma_U1}.}
\label{fig:1238}
\end{figure}

In Fig.~\ref{fig:1238}, we plot the evolution of density parameter (left panel) and the evolution of anisotropy (right panel) for an isotropic initial condition with nonvanishing velocity components~$\dot{A}^1_{1}=\dot{A}^2_{2}=\sqrt{2}\dot{A}^3_{3}=\sqrt{2}\dot{A}^8_{3}$.  
We call this initial condition isotropic because the spatial part of the energy-momentum tensor is diagonal with $T_{11}=T_{22}=T_{33}$, which is isotropic.
Note that there is no magnetic part because we have set $A^a_i=0$ at the initial time.
For $N\lesssim 25$, the anisotropy of the universe is tiny due to the cancellation of electric fields between the $z$ and $x(y)$ directions. 
In this period, the total electric density parameter~$\Omega_E$ almost coincides with that of the isotropic three-U(1) fixed point [see \eqref{Omega_multi-U1}].
Notice that the magnetic energy density is negligible in this phase.
After this stage, as the SU(2) sector grows, the density parameters associated with the SU(2) sector, $\Omega^1$ and $\Omega^3$, quickly decay, while the density parameter for the U(1) sector, $\Omega^8$, quickly converges to the value for the anisotropic U(1) case [see \eqref{Omega_U1}] in a few e-folds.
During this transient phase, the magnetic density parameter~$\Omega_B$, which is proportional to the potential,
    \be
    V_g=\bra{A^{2}_{[i} A^{3}_{j]}}^2+\bra{A^{3}_{[i} A^{1}_{j]}}^2+\bra{A^{1}_{[i} A^{2}_{j]}}^2,
    \ee 
is important.
Also, the anisotropy converges to the value for the one-U(1) case~\eqref{sigma_U1}.
Thus, the anisotropy is determined by the U(1) sector, i.e., $A^8_3$. 
Since $A^8$ has no coupling with the SU(2) sector, this state is stable.   

\subsection{A specific example: gauge-field potential with a flat direction}
\label{sec_specific}

In the previous subsection, we focused on the subgroup SU(2)\:$\otimes$\:U(1), where the anisotropy remains due to the U(1) sector. 
In this subsection, we consider another specific case where the anisotropy can survive.

Let us study the case with nonvanishing $\{A^3,A^4,A^8\}$.
We assume that only $A^3_1$, $A^4_2$, and $A^8_3$ have nontrivial initial velocities with $\dot{A}^3_{1}=\dot{A}^4_{2}=\dot{A}^8_{3}$, so that the spacetime is isotropic at the initial time.
Note that the components~$A^3_{3}$ and $A^8_{1}$ show up as the time evolves due to the nonlinear self-couplings.
This results in off-diagonal components in the metric. 
Actually, in this setup, the metric takes the form,
\be
    \D s^2=  -\D t^2 +  g_{11} \D x^2 + g_{22} \D y^2  +g_{33} \D z^2 + 2 g_{13} \D x \D z . \label{metmet}
\ee
The EOMs for the relevant components of the gauge field are given below:
\be
\begin{split}
\ddot{A}_{31} = &  - \frac{  (\sqrt{3}  A_{81} +  A_{31}) A^2_{42}
}{4 g_{22}}  
    - 2 \frac{f'}{f} \dot{\phi} \dot{A}_{31}   
    + \bra{ \fr{[g_{33},g_{11}]}{2G_{13}} - \fr{\dot{g}_{22}} {2g_{22}} } \dot{A}_{31}
    + \fr{[g_{11},g_{13}]}{G_{13}} \dot{A}_{33}, \\
\ddot{A}_{33} = &  - \frac{  (\sqrt{3}  A_{83} +  A_{33}) A^2_{42}
}{4 g_{22}}  - 2 \frac{f'}{f} \dot{\phi} \dot{A}_{33}  
     + \bra{ \fr{[g_{11},g_{33}]}{2G_{13}} - \fr{\dot{g}_{22}} {2g_{22}} } \dot{A}_{33}
     + \fr{[g_{33},g_{13}]}{G_{13}} \dot{A}_{31}, \\
\ddot{A}_{81} = &  - \frac{  (\sqrt{3}  A_{31} + 3 A_{81}) A^2_{42}
}{4 g_{22}}  - 2 \frac{f'}{f} \dot{\phi} \dot{A}_{81}  
    + \bra{ \fr{[g_{33},g_{11}]}{2G_{13}} - \fr{\dot{g}_{22}} {2g_{22}} } \dot{A}_{81} 
    + \fr{[g_{11},g_{13}]}{G_{13}} \dot{A}_{83}, \\
\ddot{A}_{83} = &  - \frac{  (\sqrt{3}  A_{33} + 3 A_{83}) A^2_{42}
}{4 g_{22}}  - 2 \frac{f'}{f} \dot{\phi} \dot{A}_{83}  
    + \bra{ \fr{[g_{11},g_{33}]}{2G_{13}} - \fr{\dot{g}_{22}} {2g_{22}} } \dot{A}_{83} 
    + \fr{[g_{33},g_{13}]}{G_{13}} \dot{A}_{81}, \\
\ddot{A}_{42} = & -\fr{A_{42}
}{4G_{13}}  \brb{g_{11}(A_{33}+\sqrt{3}A_{83})^2+g_{33}(A_{31}+\sqrt{3}A_{81})^2-2g_{13}(A_{31}+\sqrt{3}A_{81})(A_{33}+\sqrt{3}A_{83})}   \\
    &  - 2 \frac{f'}{f} \dot{\phi} \dot{A}_{42} 
    + \bra{ \fr{\dot{g}_{22}} {2g_{22}}   -  \fr{g_{11} \dot{g}_{33} +g_{33} \dot{g}_{11}  -2g_{13} \dot{g}_{13} }   {2G_{13} } } \dot{A}_{42} ,
\end{split}
\ee
where we have defined $G_{13}\coloneqq g_{11}g_{33}-g_{13}^2$ and $[f_1,f_2]\coloneqq f_1\dot{f}_2-f_2\dot{f}_1$ for any pair of functions~$f_1$ and $f_2$ of $t$.

\begin{figure}[H]
\subfigure{\includegraphics[clip, width=0.55\columnwidth]{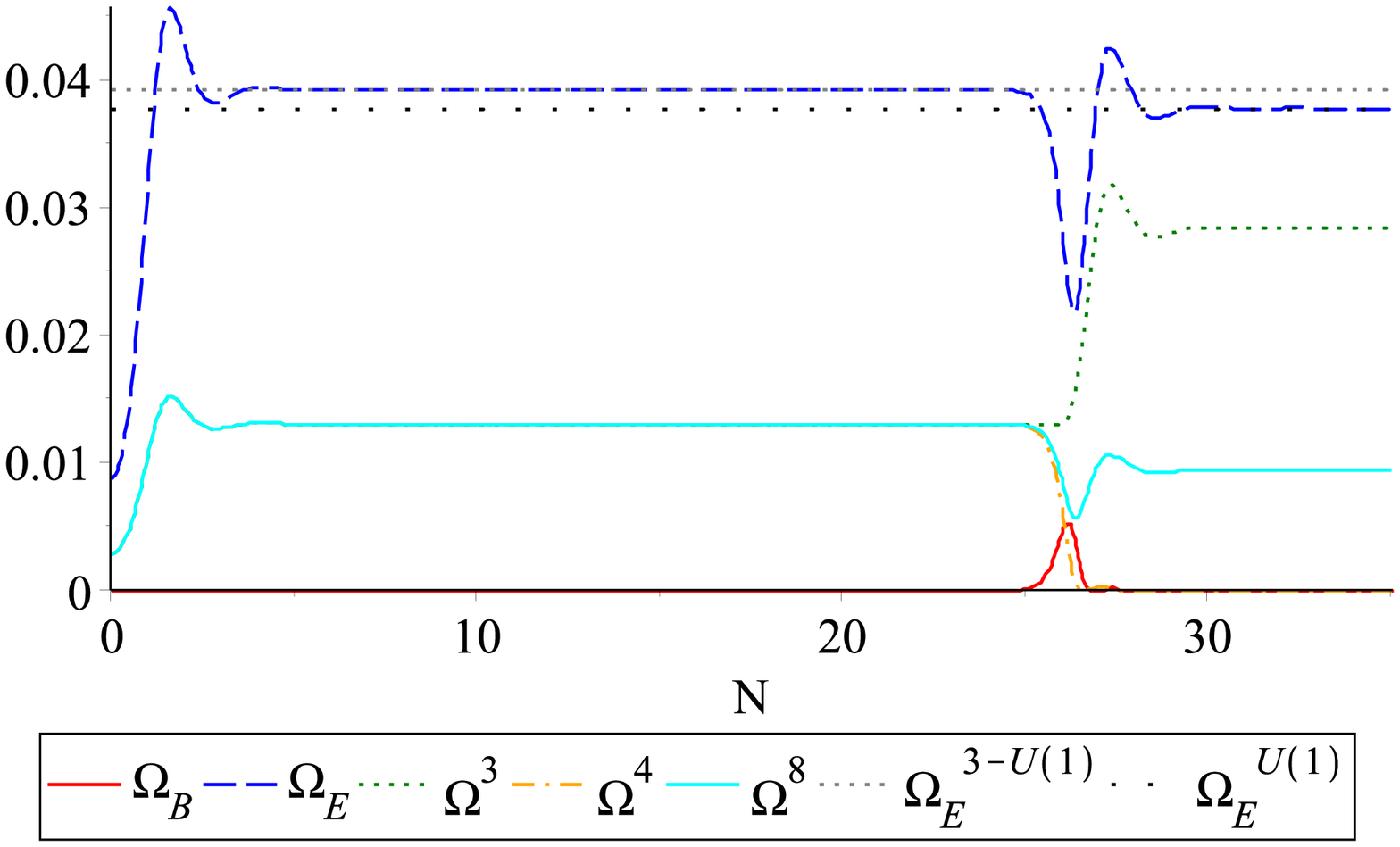}}
\subfigure{\includegraphics[clip, width=0.4\columnwidth]{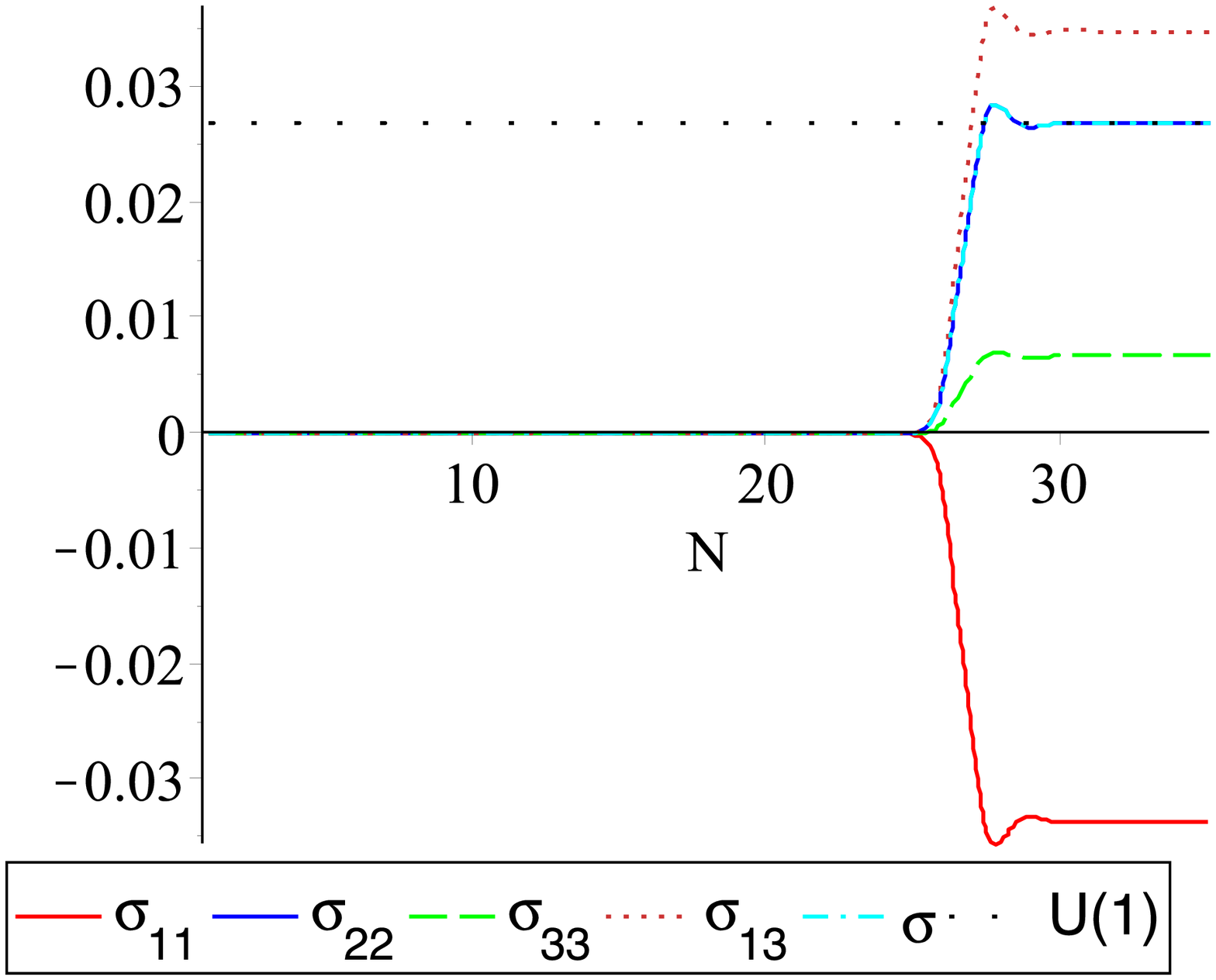}}
\vspace*{0mm}
\caption{Evolution of the density parameters of gauge-field components (left) and anisotropy (right) against e-folding number for an initial condition with $\dot{A}^3_{1}=\dot{A}^4_{2}=\dot{A}^8_{3}$. 
In the left graph, the red solid, blue dashed, green dotted, orange dash-dotted, and cyan solid curves respectively correspond to $\Omega_B$, $\Omega_E$, $\Omega^{3}$, $\Omega^{4}$, and $\Omega^{8}$.
The gray dotted and black space-dotted lines represent $\Omega_E$ for the isotropic three-U(1) case~\eqref{Omega_multi-U1} and $\Omega_E$ for the one-U(1) case~\eqref{Omega_U1}, respectively. 
In the right graph, the red solid, blue solid, green dashed, orange dotted, and cyan dash-dotted curves respectively correspond to $\sigma_{11}$, $\sigma_{22}$, $\sigma_{33}$,  $\sigma_{13}$, and the root-mean-square anisotropy~$\si$.
The black space-dotted line represents the anisotropy for the one-U(1) case~\eqref{sigma_U1}.
The curve for $\si$ almost overlaps with that of $\si_{22}$.}
\label{fig:shear348}
\end{figure}

\begin{figure}[H]
\centering
\includegraphics[clip, width=0.35\columnwidth]{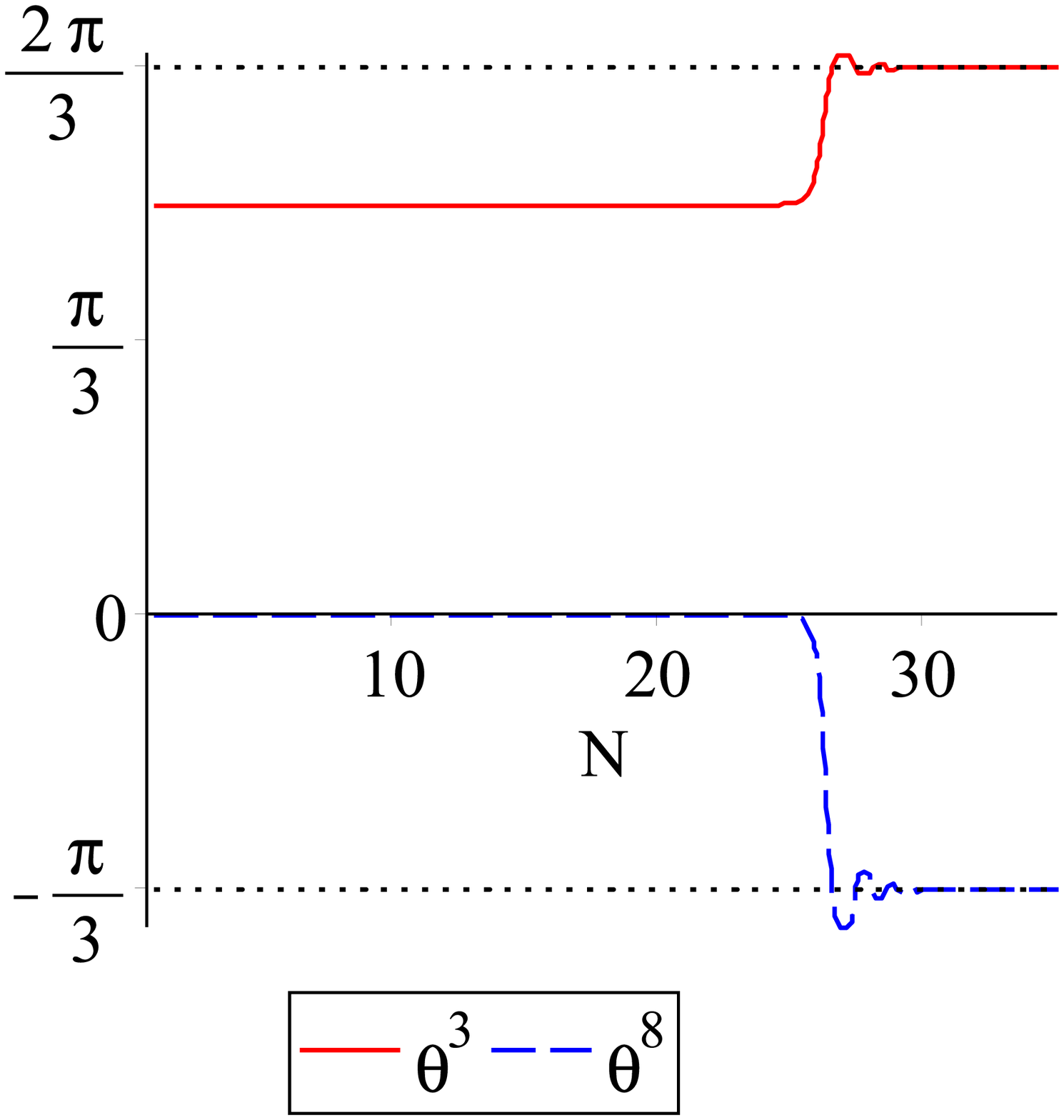}
\includegraphics[clip, width=0.35\columnwidth]{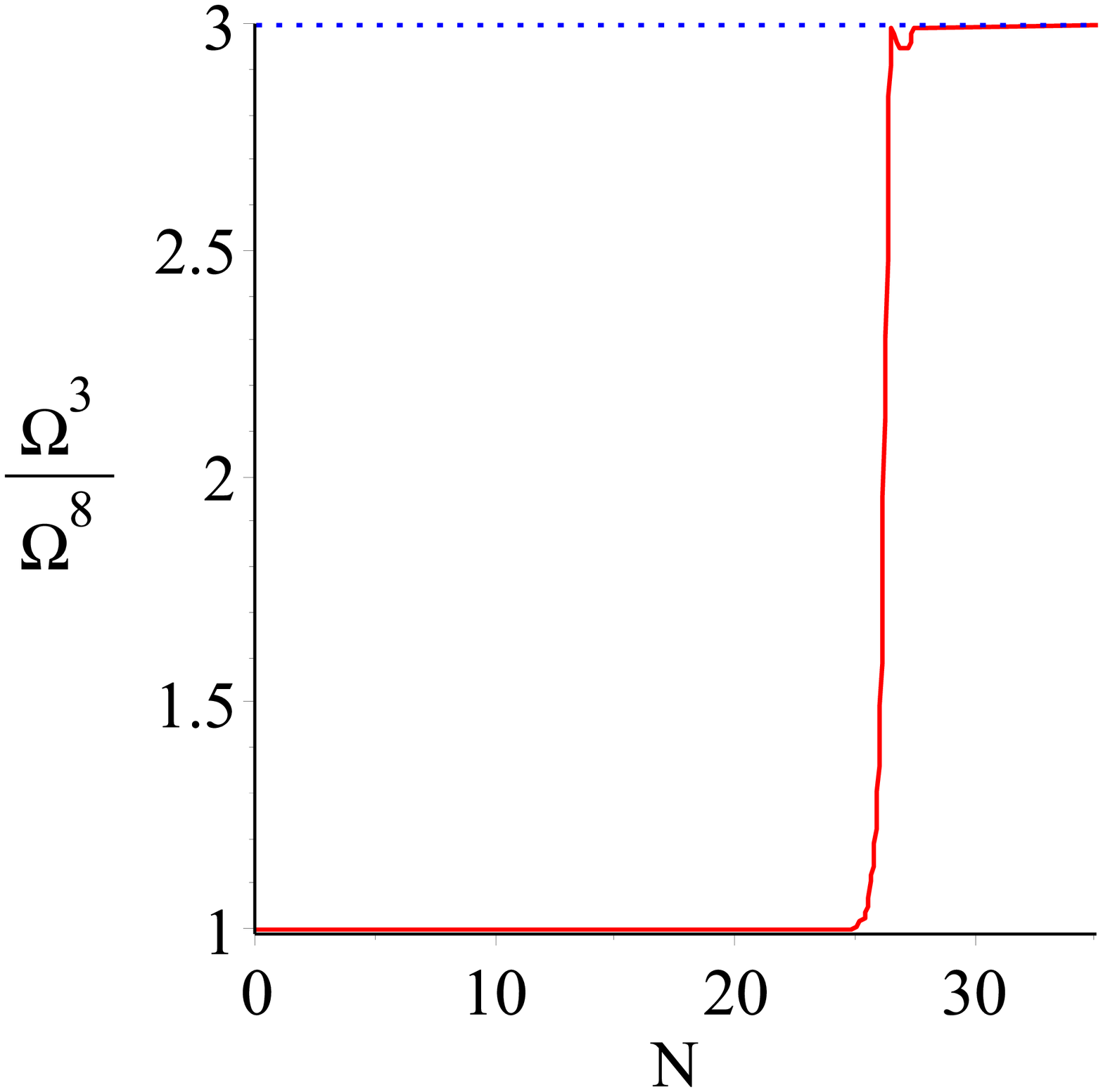}
\caption{Evolution of the gauge-field components~$A^3$ and $A^8$.
In the left panel, the red solid and blue dashed curves represent $\theta^3$ and $\theta^8$, respectively.
The right panel shows the evolution of $\Omega^3/\Omega^8$.
}
\label{fig:shear348_theta}
\end{figure}

\newpage
The evolution of the density parameters for the relevant gauge-field components and the evolution of the anisotropy are shown in Fig.~\ref{fig:shear348}. Similarly to the case of SU(2)\:$\otimes$\:U(1) subgroup in \S\ref{sec_su2*u1}, the magnetic density parameter transiently grows but then quickly decays, implying that the nonlinear self-interactions of the gauge field are important in the transient phase. 
As a result, $\Omega^4$ quickly decays after $N \sim 25$. 
However, $\Omega^3$ and $\Omega^8$ remain due to the existence of a flat direction in the potential of the gauge field~\eqref{V}.
Actually, in the present case where only $A^3$, $A^4$, and $A^8$ are nonvanishing, the potential takes the following form:
    \be
    V_g=\brb{\bra{f^{534}A^3_{[i}+f^{584}A^8_{[i}}A^4_{j]}}^2
    =\frac{1}{4}\left[\left( A^{3}_{[i}  + \sqrt{3} A^{8}_{[i} \right) A^{4}_{j]}\right]^2,
    \ee
and thus there exists a flat direction defined by
    \be
    A^3 + \sqrt{3} A^8 =0. \label{flat}
    \ee
Hence, we expect that $A^3$ and $A^8$ satisfy \eqref{flat} after the potential becomes significant. 
This is indeed the case as we show in Fig.~\ref{fig:shear348_theta}.
The left panel shows the evolution of angles~$\theta^3$ and $\theta^8$, defined by
    \be
    \sin \theta^a = \fr{\dot{A}_{a1}}{\sqrt{\dot{A}_{a1}^2+\dot{A}_{a3}^2}}, \qquad
    \cos \theta^a = \fr{\dot{A}_{a3}}{\sqrt{\dot{A}_{a1}^2+\dot{A}_{a3}^2}},
    \ee
for $a=3,8$.
We see that $\dot{A}^3$ and $\dot{A}^8$ are anti-parallel after a sufficiently long time.
The right panel shows the evolution of the ratio of $\Omega^3$ to $\Omega^8$, from which we see that $\Omega^3/\Omega^8\to 3$ at late times.
Combining these results, we find that $\dot{A}^3=-\sqrt{3}\dot{A}^8$ at late times, which is consistent with \eqref{flat}.
Since the gauge field is trapped in the flat direction~\eqref{flat}, the dynamics is similar to the one-U(1) case studied in \S\ref{ssec_U(1)}.
Indeed, as shown in Fig.~\ref{fig:shear348}, the total electric density parameter~$\Omega_E$ and the anisotropy~$\si$ approach to the values for the one-U(1) case.
We note that, although we have only two nonvanishing components of the gauge field (i.e., $A^3$ and $A^8$) at late times, the final state here is different from the stable fixed point for the two-U(1) case studied in \S\ref{ssec_2U(1)}, where the two U(1) gauge fields are orthogonal to each other.
This is due to the existence of the flat direction in the potential of the gauge field mentioned above.

Likewise, for the case with nonvanishing $\{A^3,A^5,A^8\}$, we find the same flat direction as \eqref{flat}.
On the other hand, for the cases with nonvanishing $\{A^3,A^6,A^8\}$ or $\{A^3,A^7,A^8\}$, we find the following flat direction:
    \be
    A^3-\sqrt{3}A^8=0. \label{flat2}
    \ee
As we shall discuss in \S\ref{sec:Extension}, the existence of such flat directions is a clear difference of SU(3) from SU(2).

\subsection{General cases}
\label{sec_general}

\begin{figure}[H]
\centering
\subfigure{\includegraphics[clip, width=0.3\columnwidth]{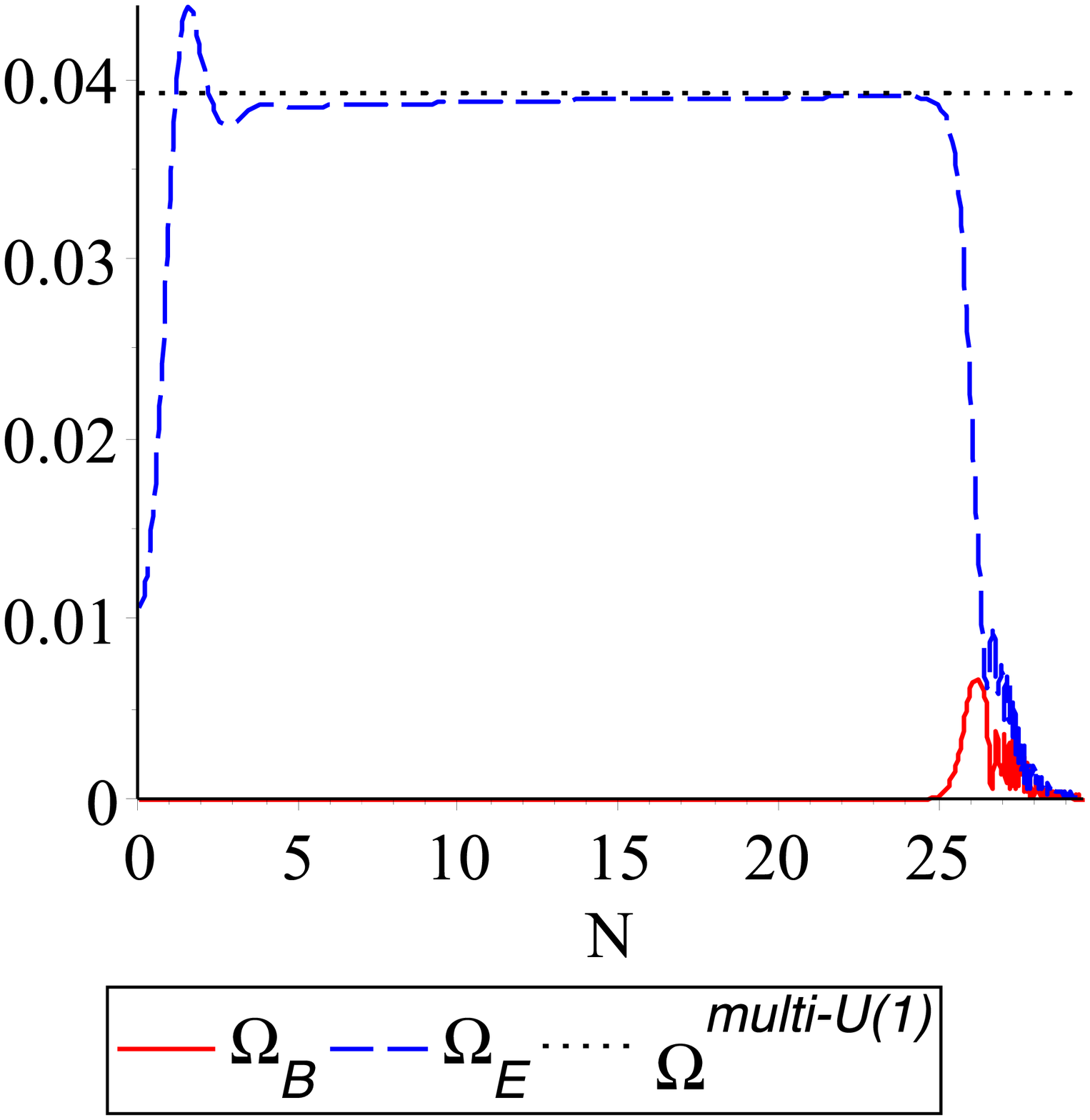}}
\subfigure{\includegraphics[clip, width=0.35\columnwidth]{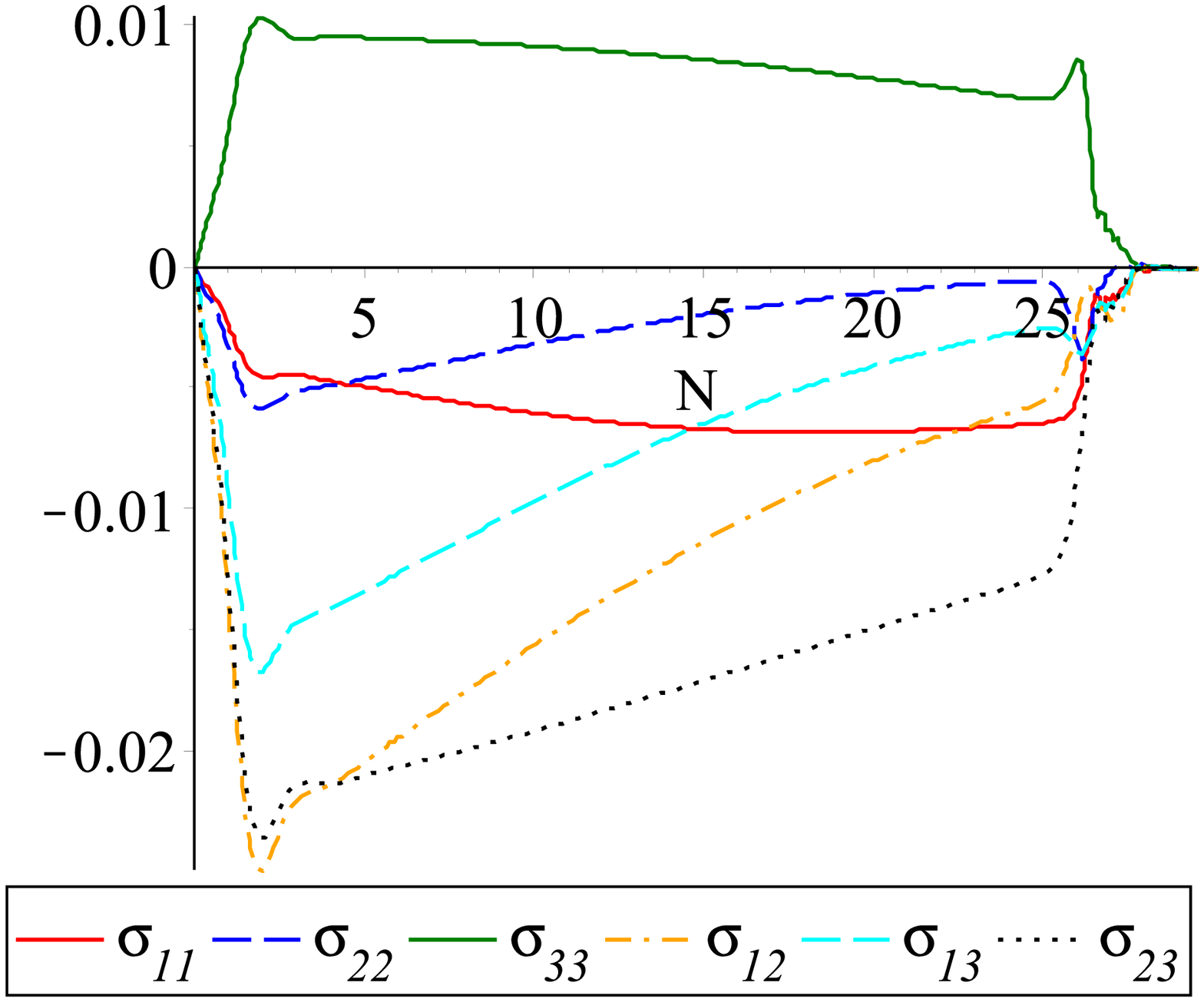}}
\subfigure{\includegraphics[clip, width=0.3\columnwidth]{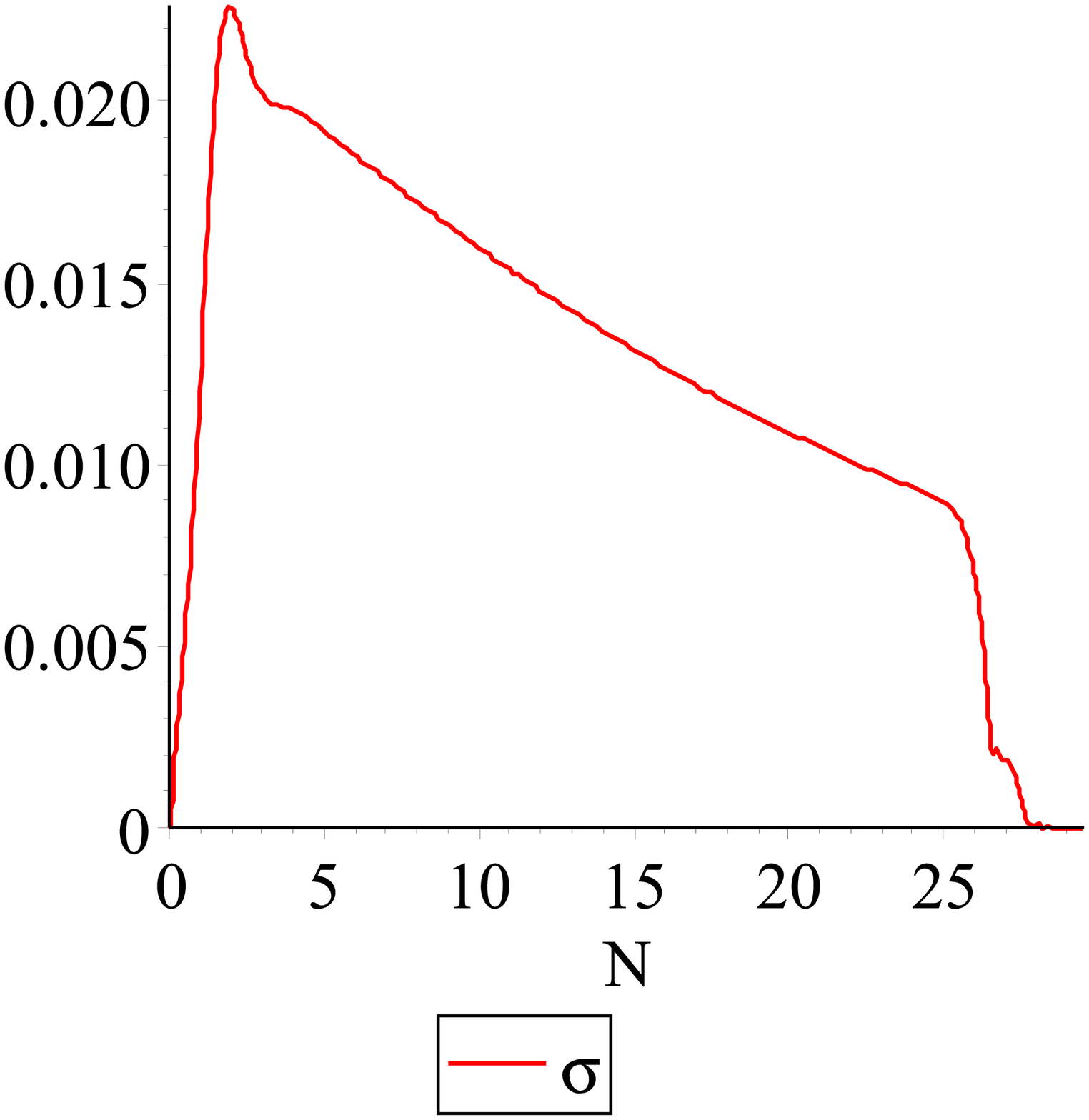}}
\caption{Evolution of the density parameters (left), $\sigma_{ij}$ (middle), and $\sigma$ (right) against e-folding number for an initial condition with all $\dot{A}^a_{i}$'s having the same order. 
In the left graph, the red solid and blue dashed curves correspond to the total electric and magnetic density parameters, respectively.
The black dotted line represents the total electric density parameter for the case of isotropic multi-U(1) gauge fields~\eqref{Omega_multi-U1}. 
In the middle graph, the red solid, blue dashed, green solid, orange dash-dotted, cyan dashed, and black dotted curves correspond to $\sigma_{11}$, $\sigma_{22}$, $\sigma_{33}$, $\sigma_{12}$, $\sigma_{13}$, and $\sigma_{23}$, respectively.
}
\label{fig:all}
\end{figure}

So far, we have considered the special cases where the anisotropy survives. 
However, in general, the anisotropy decays once the nonlinearity of gauge fields becomes important.
To see this, we study the situation where all the gauge-field components have initial velocities of the same order.
In Fig.~\ref{fig:all}, we show the evolution of the density parameters for the gauge field and the spacetime anisotropy.
The anisotropic expansion of spacetime lasts until $N\sim 25$, and then the nonlinear self-couplings of the gauge field become important and the anisotropy decays.
Indeed, the magnetic density parameter~$\Omega_B$, which measures the effect of nonlinear self-couplings, is comparable to the total electric density parameter~$\Omega_E$ at around $N\sim 25$.

To reiterate, the anisotropy decays at late times unless we fine-tune the initial condition as in \S\ref{sec_su2*u1} and \S\ref{sec_specific}. 
Thus, the cosmic no-hair conjecture generically holds.
In a realistic universe, it is reasonable to expect that all the components of the SU(3) gauge field have initial values of the same order, and hence the expansion of the universe should become isotropic after a sufficiently long time after the onset of inflation.

\section{More on Inflation with Non-Abelian Gauge Fields} \label{sec:Extension}

We have studied inflationary universes with an SU(3) gauge field.
One can generalize the discussion to a non-Abelian SU($N$) gauge field for arbitrary $N$.
In this section, we will show that there are flat directions in the potential of an SU($N$) gauge field if $N \geq 3$.

Let us consider an SU($N$) gauge field with $N\ge 2$. 
The elements of the Cartan subalgebra
are represented by
\begin{eqnarray}
(H_m )_{ij} 
= \frac{1}{\sqrt{2m(m+1)}}\left(
\sum^m_{k=1} \delta_{ik} \delta_{jk} -m\delta_{i,m+1}\delta_{j,m+1}\right)
, \qquad m=1,2,\cdots , N-1 .
\end{eqnarray}
In particular, for $N=3$, we have $H_1=T^3$ and $H_2=T^8$.
The algebra is completely determined by the following simple roots~\cite{Georgi:1999wka}:
\be
\begin{split}
\alpha^1 &= \left( 1, 0 , \cdots  ,0  \right), \\
\alpha^2 &= \left( -\frac{1}{2}, \frac{\sqrt{3}}{2} , 0, \cdots , 0  \right),\\
\alpha^3 &= \left( 0, -\frac{1}{\sqrt{3}} , \sqrt{\frac{2}{3}}, 0 , \cdots , 0 \right),\\
&\vdots \\ 
\alpha^m &=  \left(  0,  \cdots, 0, -\sqrt{\fr{m-1}{2m}} , \sqrt{\fr{m+1}{2m}},0,\cdots,0  \right)   ,    \\
&\vdots \\ 
\alpha^{N-1} &= \left( 0, \cdots 0, -\sqrt{\frac{N-2}{2(N-1)}} ,
\sqrt{\frac{N}{2(N-1)}}\right) ,
\end{split}
\ee
where each root is an $(N-1)$-dimensional vector.
From these, one can find flat directions in the potential of the gauge field.
As discussed in \S\ref{sec_specific}, flat directions stem from the coexistence
of nonzero structure constants like $f^{acd}$ and $f^{bcd}$ ($a \neq b$).
This happens when we have simple roots of $\alpha^2 , \alpha^3 , \dots , \alpha^{N-1}$, so that there is no flat direction in the SU(2) case.
In this sense, SU(2) is clearly different from other non-Abelian gauge groups.

More concretely, for $N\ge 3$, there exists a gauge-field component~$A^n$ for each $m=2,3,\cdots,N-1$ such that the gauge-field potential contains the following combination:
    \be
    \bra{A^{m-1}_{[i}-\sqrt{\fr{m+1}{m-1}}A^m_{[i}}A^n_{j]},
    \ee
and hence there are flat directions defined by
\begin{eqnarray}
  A^{m-1}-\sqrt{\fr{m+1}{m-1}}A^m=0. \label{flat_SU(N)}
\end{eqnarray} 
Namely, for each $m$, the value of the potential remains unchanged under any change of $A^{m-1}$ and $A^m$ satisfying \eqref{flat_SU(N)} so long as the motion of the gauge field is (at least dynamically) constrained within the hypersurface in the configuration space spanned by $A^{m-1}$, $A^m$, and $A^n$.
Specifically, for $N=3$, the flat direction is given by
    \be
    A^1-\sqrt{3}A^2=0.
    \ee
Note that $A^m$ in this section is the gauge-field component associated with the element~$H_m$ of the Cartan subalgebra, i.e., $A^m=2\,{\rm Tr}\,(\mathbf{A}H_m)$, and hence should be distinguished from $A^a$ in \S\ref{sec:SU3}, where we expanded the gauge field in terms of the SU(3) generators~$T^a$.
As mentioned earlier, since $H_1=T^3$ and $H_2=T^8$, $A^1$ and $A^2$ here should be identified as $A^3$ and $A^8$ in \S\ref{sec:SU3}, respectively.
Hence, the above flat direction is nothing but the one \eqref{flat2}.

Furthermore, there are other flat directions corresponding to positive roots other than the simple roots.
For example, in the case of SU(3), a positive root $(1/2 , \sqrt{3}/2)$ indicates another flat direction: 
\be
    A^1 + \sqrt{3}A^2=0.
\ee
Indeed, this is the flat direction of \eqref{flat}.
Similarly, for $N \geq 4$, positive roots other than the simple roots give rise to additional flat directions. 
Although we focused on the SU($N$) group in this section, 
the above argument can be generalized to other Lie groups.

\section{Conclusion} \label{sec:conclusion}

We studied inflationary universes in the presence of an SU(3) gauge field. 
We numerically solved the system of coupled EOMs to obtain the time evolution of the spacetime, the inflaton, and the gauge field.
In general, even if we start from an isotropic spacetime, the anisotropy can be generated if the gauge field has an initial velocity.

There are special cases where the generated anisotropy does not decay and there remains a finite anisotropy.
As an example, in \S\ref{sec_su2*u1}, we studied the situation where the components of the SU(3) gauge field can be separated into the SU(2) and U(1) sectors.
The energy density of the SU(2) sector decays due to the nonlinear self-interactions, but that of the U(1) sector remains, and hence the anisotropic expansion of spacetime lasts.
The resultant anisotropy coincides with the one obtained in \cite{Kanno:2010nr}, where an exact solution of power-law anisotropic inflation with a U(1) gauge field was studied.
We studied another interesting case in \S\ref{sec_specific}, where the three components~$A^3_1$, $A^4_2$, and $A^8_3$ have the same initial velocity.
In this case, after a sufficiently long time, $A^4$ becomes negligible while $A^3$ and $A^8$ survive with $A^3+\sqrt{3}A^8=0$.
This phenomenon arises due to flat directions in the potential of the gauge field. 
Namely, the gauge field is trapped in the flat direction defined by $A^3+\sqrt{3}A^8=0$ and behaves in a similar manner to the case of a U(1) gauge field.
Also, as we clarified in \S\ref{sec:Extension}, such flat directions exist in general for non-Abelian gauge fields whose associated Lie group has a rank higher than one. 
It should be noted that there is no flat direction in the potential for an SU(2) gauge field.
This gives rise to an interesting inflationary scenario
with an SU(3) gauge field, which cannot be realized in the SU(2) case.

On the contrary, in a realistic universe, it is reasonable to expect that all the components of the SU(3) gauge field have nonvanishing initial values of the same order of magnitude. 
We considered such a situation in \S\ref{sec_general}.
We found that the generated anisotropy eventually decays due to the nonlinear self-couplings of the gauge field (see also an 
analogous result in \cite{Maleknejad:2011jr}).
In this sense, the cosmic no-hair conjecture holds. 
However, the transient anisotropy should exist practically on the large scales and its effect would be imprinted on the cosmic microwave background and the large-scale structure.

There are several interesting directions for further developments. 
In this paper, we have considered inflation with an SU(3) gauge field as a first step. 
It would be intriguing to study general non-Abelian gauge fields such as SU($N$) in detail. 
It is also interesting to investigate the Schwinger effect in the presence of non-Abelian gauge fields.
Studying the Chern-Simons--type interaction for an SU(3) gauge field instead of the gauge kinetic function may also
give interesting features. 
Another possible extension would be to study models with multiple scalar fields, where the field-space metric is not necessarily flat.
Then, the nontrivial kinetic structure may change the dynamics~\cite{Chen:2021nkf}.
Thus, it is worth studying the cosmic no-hair conjecture in a more general context.
We leave these issues for future study.

\acknowledgments{
P.\,G.\ was supported by Japanese Government (MEXT) Scholarship and China Scholarship Council.
K.\,T.\ was supported by JSPS (Japan Society for the Promotion of Science) KAKENHI Grant Number JP21J00695. 
A.\,I.\ was supported by JSPS KAKENHI Grant Number JP21J00162.
J.\,S.\ was in part supported by JSPS KAKENHI Grant Numbers JP17H02894, JP17K18778, JP20H01902.
}

\renewcommand{\theequation}{A\arabic{equation}}
\setcounter{equation}{0}
\section*{\texorpdfstring{A\lowercase{ppendix}: SU(3) \lowercase{gauge field in the axially symmetric} B\lowercase{ianchi type} I \lowercase{spacetime}}{SU(3) gauge field in the axially symmetric Bianchi type I spacetime}}
\label{Appendix}

In this appendix, we derive possible configurations of an SU(3) gauge field in the axially symmetric Bianchi type~I spacetime.
To this end, we extend the discussion for the case of an SU(2) gauge field~\cite{Murata:2011wv,Darian:1996jf}
to an SU(3) gauge field.

First of all, from the translation and the local SU(3) gauge invariance, one can write an SU(3) gauge field as
\begin{equation}
  A^{a} = P^{a}(t) \D x + Q^{a}(t) \D y + R^{a}(t) \D z.
\end{equation}
In addition, we impose the axial symmetry along a particular direction, say, the $z$-direction on it.
The rotational transformation along the $z$-direction, which is generated by a killing vector~$\xi = x \partial_{y} - y\partial_{x}$, is given by
\begin{equation}
  \mathcal{L}_{\xi} A^{a} = Q^{a}(t) \D x - P^{a}(t) \D y . \label{rot}
\end{equation}
In order to preserve the rotational symmetry,
\eqref{rot} must be absorbed by the residual global SU(3) transformation:
\begin{equation}
  \delta A^{a} = i[A, u]^{a} = f^{abc} u^{b} \brb{ P^{c}(t) \D x + Q^{c}(t) \D y + R^{c}(t) \D z } ,
\end{equation}
where $u^{a}$'s are constant.
Therefore, we require
\begin{equation}
  \mathcal{L}_{\xi} A^{a} = \delta A^{a} .  \label{ab}
\end{equation}
Configurations of $A^{a}$ which satisfy this relation can be classified according to the direction and amplitude of $u^{a}$.
A trivial case is $u^{a}=0$, we have the condition~$P^{a}(t) = Q^{a}(t) =0$.
Let us consider cases of $u^{a} = u^{3},u^{4},u^{8}$ as representative examples.
We first consider the case of $u^{a} = u^{3}$.
In this case, \eqref{ab} admits nontrivial configurations of $A^{a}$ only if $u^{3}=\pm 1$ or $\pm 2$.
For instance, $u^{3}=1$ yields
\begin{equation}
\left\{ \,
    \begin{aligned}
    & P(t) = P^{1}(t)T^{1} + P^{2}(t)T^{2} , \\
    & Q(t) = -P^{2}(t)T^{1} + P^{1}(t)T^{2} ,  \\
    & R(t) = R^{3}(t)T^{3} + R^{8}(t)T^{8},   \label{su2su1}
    \end{aligned}
\right. 
\end{equation}
and $u^{3}=2$ gives
    \begin{equation}
\left\{ \,
    \begin{aligned}
    & P(t) = P^{4}(t)T^{4} + P^{5}(t)T^{5} + P^{6}(t)T^{6} + P^{7}(t)T^{7}, \\
    & Q(t) = -P^{5}(t)T^{4} + P^{4}(t)T^{5} + P^{7}(t)T^{6} - P^{6}(t)T^{7},  \\
    & R(t) = R^{3}(t)T^{3} + R^{8}(t)T^{8}.
    \end{aligned}
\right.
\end{equation}
The case of \eqref{su2su1} includes the SU(2)\:$\otimes$\:U(1) subgroup we studied in \S\ref{sec_su2*u1}.
Next, in the case of $u^{a} = u^{4}$, we have a solution of \eqref{ab} only if $u^{4} = \pm 1$ or $\pm 2$. 
For $u^{4} = 1$, we have
\begin{equation}
\left\{ \,
    \begin{aligned}
    & P(t) = P^{5}(t)T^{5} + P^{3}(t)\bra{T^{3}+\sqrt{3}T^{8}}, \\
    & Q(t) = -2P^{3}(t)T^{5} + \fr{1}{2}P^{5}(t)\bra{T^{3}+\sqrt{3}T^{8}},  \\
    & R(t) = R^{4}(t)T^{4} + R^{8}(t)\bra{-\sqrt{3}T^{3}+T^{8}},
    \end{aligned}
\right.
\end{equation}
while, for $u^{4}=2$, we obtain 
\begin{equation}
\left\{ \,
    \begin{aligned}
    & P(t) = P^{1}(t)T^{1} + P^{2}(t)T^{2} + P^{6}(t)T^{6} + P^{7}(t)T^{7}, \\
    & Q(t) = P^{7}(t)T^{1} + P^{6}(t)T^{2} - P^{2}(t)T^{6} - P^{1}(t)T^{7},  \\
    & R(t) = R^{4}(t)T^{4} + R^{8}(t)\bra{-\sqrt{3}T^{3}+T^{8}}.
    \end{aligned}
\right.
\end{equation}
Finally, when $u^{a} = u^{8}$, the only possibility is $u^{8} = \pm 2/\sqrt{3}$.
For $u^{8} = 2/\sqrt{3}$, the configuration of the gauge field which satisfies \eqref{ab} is 
\begin{equation}
\left\{ \,
    \begin{aligned}
    & P(t) = P^{4}(t)T^{4} + P^{5}(t)T^{5} + P^{6}(t)T^{6} + P^{7}(t)T^{7}, \\
    & Q(t) = -P^{5}(t)T^{4} + P^{4}(t)T^{5} - P^{7}(t)T^{6} + P^{6}(t)T^{7},  \\
    & R(t) = R^{1}(t)T^{1} + R^{2}(t)T^{2} + R^{3}(t)T^{3} + R^{8}(t)T^{8} .
    \end{aligned}
\right.
\end{equation} 

In practice, it is necessary to impose the Yang-Mills constraint, i.e., 
    \be
    \nabla_{i} F^{ai0} + 
    f^{abc} A^b_{i} F^{ci0} = 0, \label{YMc}
    \ee
which is nothing but the time component of the EOMs for the gauge field~\eqref{EOMs_gauge}. 
This further constrains the gauge-field configuration.
More explicitly, \eqref{YMc} can be reduced as
\begin{equation}
  f^{abc}\left[  \left( P^b(t) \dot{P}^c(t) + Q^b(t)\dot{Q}^c(t) \right)g^{11}(t)  
                 + R^b(t) \dot{R}^c(t) g^{33}(t)  \right] = 0 ,
\end{equation} 
in the axially symmetric Bianchi type I spacetime.
For instance, for \eqref{su2su1}, the above constraint yields $P^2/P^1={\rm const}$.
Likewise, one can obtain some relations among the functions~$P^a(t)$, $Q^a(t)$, and $R^a(t)$ for other cases.


\bibliographystyle{mybibstyle}
\bibliography{bib}

\end{document}